\documentclass[a4paper,12pt]{article}

\usepackage[margin=2.8cm]{geometry} 

\usepackage{graphicx}
\usepackage{amssymb}
\usepackage{amsmath}
\usepackage{amsfonts}
\usepackage{amsthm}

\usepackage{subfigure}
\usepackage{epstopdf}
\usepackage{appendix}

\newcommand{\N}{\mathbb{N}}
\newcommand{\R}{\mathbb{R}}

\newcommand{\zM}{z_{max}}
\newcommand{\zm}{z_{min}}
\newcommand{\zf}{z_{F}}
\newcommand{\zw}{z_{W}}
\newcommand{\zV}{z_{vax}}

\title{Mathematical models for vaccination, waning immunity and immune system boosting:\\ a general framework}
\author{M.~V. Barbarossa \thanks{Bolyai Institute,
University of Szeged, H-6720 Szeged, Aradi v\'{e}rtan\'{u}k tere 1, Hungary,
\textit{barbaros@math.u-szeged.hu}} \, and 
      G.~R\"ost \thanks{Bolyai Institute,
University of Szeged, H-6720 Szeged, Aradi v\'{e}rtan\'{u}k tere 1, Hungary,
\textit{rost@math.u-szeged.hu}}}
\date{}

\begin{document}

\maketitle

\begin{abstract}
When the body gets infected by a pathogen or receives a vaccine dose, the immune system develops pathogen-specific immunity. Induced immunity decays in time and years after recovery/vaccination the host might become susceptible again. Exposure to the pathogen in the environment boosts the immune system thus prolonging the duration of the protection. Such an interplay of within host and population level dynamics poses significant challenges in rigorous mathematical modeling of immuno-epidemiology.
The aim of this paper is twofold. First, we provide an overview of existing models for waning of disease/vaccine-induced immunity and immune system boosting. Then a new modeling approach is proposed for SIRVS dynamics, monitoring the immune status of individuals and including both waning immunity and immune system boosting. We show that some previous models can be considered as special cases or approximations of our framework.\\
\ \\
\textit{KEYWORDS:} {Immuno-epidemiology;\, Waning immunity;\, Immune status;\, Boosting;\, Physiological structure;\, Reinfection;\, Delay equations;\, Vaccination}\\
\ \\
\textit{AMS Classification:} {92D30;\, 35Q91;\, 34K17}
\end{abstract}

\section{Introduction}
Models of SIRS type are a traditional topic in mathematical epidemiology. Classical approaches present a population divided into susceptibles (S), infectives (I) and recovered (R), and consider interactions and transitions among these compartments \cite{Brauer2001}. Susceptibles are those hosts who either did not contract the disease in the past or lost immunity against the disease-causing pathogen. When a susceptible host gets in contact with an infective one, the pathogen can be transmitted from the infective to the susceptible and with a certain probability the susceptible host becomes infective himself. After pathogen clearance the infective host recovers and becomes immune for some time, afterward he possibly becomes susceptible again (in certain cases one can talk of life-long immunity). The model can be extended by adding vaccination. Vaccinees (V) are protected from infection for some time, usually shorter than naturally infected hosts.\\
\ \\
From the in-host point of view, immunity to a pathogen is the result of either active or passive immunization. The latter is a transient protection due to the transmission of antibodies from the mother to the fetus through the placenta. The newborn is thus immune for several months after birth \cite{McLean1988a}. Active immunization is either induced by natural infection or can be achieved by vaccine administration \cite{Siegrist2008,KubyImmBook}.

\begin{figure}[!]
\centering
\includegraphics[width=0.9\columnwidth]{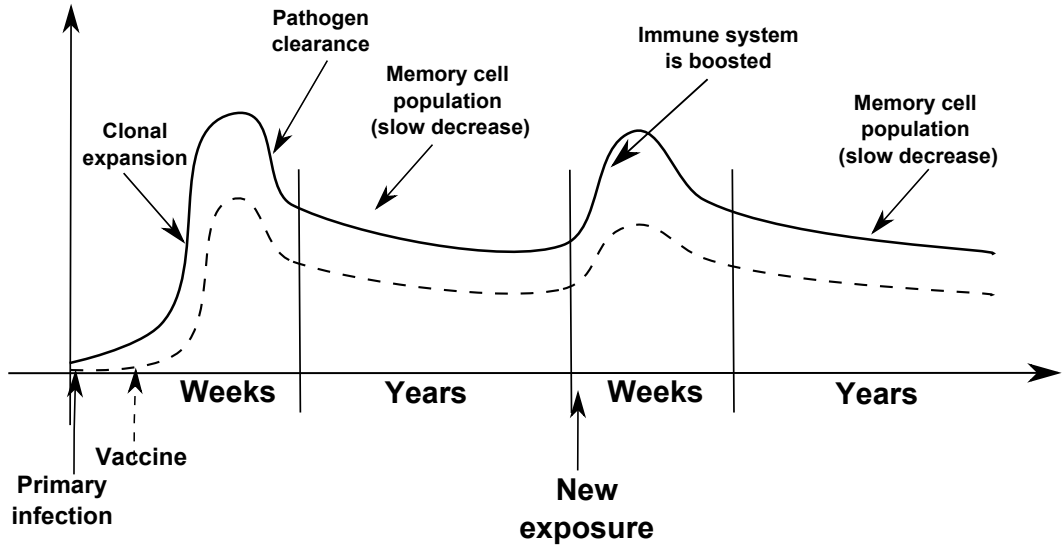}
\caption{Level of pathogen-specific immune cells with respect to the time. The solid line represents the case of natural infection, the dotted line represents the immune status of a vaccinated host. Generation of memory cells takes a few weeks: once primary infection (respectively, vaccination) occurred, the adaptive immune system produces a high number of specific immune cells (clonal expansion). After pathogen clearance, specific immune cells (memory cells) are maintained for years at a level that is much higher than in a naive host. Memory cells are activated in case of secondary infection.}
\label{Fig:introfig1}
\end{figure}

\indent Let us first consider the case of natural infection. A susceptible host, also called \textit{naive host}, has a very low level of specific immune cells for a pathogen (mostly a virus or a bacterium). 
The first response to a pathogen is nonspecific, as the innate immune system cannot recognize the physical structure of the pathogen. The innate immune response slows down the initial growth of the pathogen, while the adaptive (pathogen-specific) immune response is activated. Clonal expansion of specific immune cells (mostly antibodies or CTL cells) and pathogen clearance follow. The population of pathogen-specific immune cells is maintained for long time at a level that is much higher than in a naive host. These are the so-called \textit{memory cells} and are activated in case of secondary infection (see Figure \ref{Fig:introfig1}, adapted from \cite{BarbarossaRostJoMB}.). Memory cells rapidly activate the immune response and the host mostly shows mild or no symptoms \cite{Antia2005}.

\indent Each exposure to the pathogen might have a boosting effect on the population of specific memory cells. Indeed, the immune system reacts to a new exposure as it did during primary infection, thus yielding an increased level of memory cells. Though persisting for long time after pathogen clearance, the memory cell population slowly decays and in the long run the host might lose his pathogen-specific immunity \cite{Wodarz2007}.

\indent  Vaccine-induced immunity works in a similar way as immunity induced by the natural infection. Agents contained in 
vaccines resemble, in a weaker form, the disease-causing pathogen and force a specific immune reaction without leading to the disease. If the vaccine is successful, the host is immunized for some time. Vaccinees experience immune system boosting and waning immunity, just as hosts recovered from natural infection do. In general, however, disease-induced immunity induces a much longer lasting protection than vaccine-induced immunity does \cite{Siegrist2008}.

\noindent Waning immunity might be one of the factors which cause, also in highly developed regions, recurrent outbreaks of infectious diseases such as measles, chickenpox and pertussis. On the other side, immune system boosting due to contact with infectives prolongs the protection duration. In a highly vaccinated population there are a lot of individuals with vaccine-induced immunity and few infection cases, as well as many individuals with low level of immunity. In other words, if a high portion of the population gets the vaccine, there are very few chances for exposure to the pathogen and consequently for immune system boosting in protected individuals.\\
\ \\
In order to understand the role played by waning immunity and immune system boosting in epidemic outbreaks, in the recent past several mathematical models were proposed. Few of these models describe only in-host processes during and after the infection \cite{Wodarz2007,Heffernan2008}. Many more models, formulated in terms of ordinary differential equations (ODEs), consider the problem only at population level, defining compartments for individuals with different levels of immunity and introducing transitions between these compartments \cite{Dafilis2012,Heffernan2009}. Vaccinated hosts or newborns with passive immunity are often included in the model equations and waning of vaccine-induced or passive immunity are observed \cite{Rouderfer1994,Mossong1999,Glass2003b,Grenfell2012,Lavine2011,Arino2006,Mossong2003}. 

\indent To describe the sole waning immunity process, authors have sometimes chosen delay differential equation (DDE) models with constant or distributed delays \cite{Kyrychko2005,Taylor2009,Blyuss2010,Bhat2012,Belair2013}. The delay represents the average duration of the disease-induced immunity. However, neither a constant nor a distributed delay allows for the description of immune system boosting.

\indent Models which include partial differential equations (PDEs) mostly describe an age-structured population \cite{mclean1988,Katzmann1984,Rouderfer1994} and consider pathogen transmission among the different age groups (newborns, children, pupils, adults, \ldots). Rare examples suggest a physiologically structured approach with populations structured by the level of immunity, coupling within-host and between-hosts dynamics \cite{Martcheva2006,BarbarossaRostJoMB}.\\
\ \\
The goal of the present book chapter is twofold. On the one side, we found necessary to provide a comprehensive overview of previously published models for waning of disease/vaccine-induced immunity and immune system boosting (Sect.~ \ref{sec:overview}). On the other side, in Sect.~\ref{sec:framework} we propose a new modeling framework for SIRVS dynamics, monitoring the immune status of individuals and including both waning immunity and immune system boosting.

\section{Mathematical models for waning immunity and immune system boosting}
\label{sec:overview}
In the following we provide an overview on previous mathematical models for waning immunity and immune
system boosting. We shall classify these models according to their mathematical structure (systems of ODEs, PDEs or DDEs).

\subsection{Systems of ODEs}
Mossong and coauthors were among the first to suggest the inclusion of individuals with waning immunity in classical SIRS systems \cite{Mossong1999}. Motivated by the observation that measles epidemics can occur even in highly vaccinated populations, the authors set up a model to study the waning of vaccine-induced immunity and failure of seroconversion as possible causes for recurrent outbreaks. Their compartmental model includes hosts with the so-called ``vaccine-modified measles infection'' (VMMI) which can occur in people with some degree of passive immunity to the virus, including those previously vaccinated. Assuming that not all vaccinees are protected from developing VMMI, the authors classify vaccinees into three groups: immediately susceptible to VMMI (weak response), temporarily protected who become susceptible to VMMI due to waning of vaccine-induced immunity (intermediate response), and permanently protected from VMMI (strong response). Infection occurs due to contact with infectious individuals (both regular measles infection and VMMI). The resulting compartmental model includes waning of vaccine-induced immunity but not of disease-induced immunity, nor immune system boosting. Similar to McLean and Blower \cite{McLean1993}, Mossong et~al. define a parameter $\phi$ to describe the impact of the vaccine: if $\phi<1$, then vaccine failure is possible. Analytical results in \cite{Mossong1999} show that the main effect of VMMI is to increase the overall reproduction number of the infection.\\
\ \\
Inspired by Mossong's work, in 2003-2004 Glass, Grenfell and coauthors \cite{Glass2003,Glass2004b,Glass2004} proposed modifications and extensions of the system in \cite{Mossong1999}. The basic model is similar to the ODE system in \cite{Mossong1999}, with a group of subclinical cases which carry the pathogen without showing symptoms \cite{Glass2003b}. In addition, the distribution of antibody levels in immune hosts (included in the ODEs coefficients) and immune system boosting are introduced: the average antibody level in an immune host increases due to contact with infective or subclinical hosts. This model was used to fit measles data in England \cite{Glass2004}. In \cite{Glass2004b} the basic model was extended to consider measles transmission in a meta-population with $N$ patches.\\
\indent Immune system boosting in vaccinees was further studied in \cite{Grenfell2012}. In this work two models are introduced. In the first one vaccinees are separated from non-vaccinated hosts. Both groups of individuals are classified into susceptible, infective and immune, but in contrast to the models in \cite{Glass2003,Glass2004b,Glass2004,Mossong1999}, there is no compartment for subclinical cases. Non-vaccinated hosts do not undergo immune system boosting. For vaccinated hosts the authors include a so called ``self-boosting'' of vaccine, so that contact with infectives moves susceptible vaccinees to the immune vaccinated compartment. The second model extends the first one with a new compartment for hosts with waning immunity (W). These can receive immune system boosting due to contact with infectives or move back to the susceptible compartment due to immunity loss. Numerical simulations show possible sustained oscillations. The SIRWS system was partially analyzed by Dafilis et~al. \cite{Dafilis2012}.\\
\ \\
Heffernan and Keeling \cite{Heffernan2008} proposed an in-host model to understand the behavior of the immune system during and after an infection. Activation of immune system effectors and production of memory cells depend on the virus load. When not stimulated by the virus, the number of activated cells decays (waning immunity). Vaccination is simulated by changing the initial conditions for the virus load. Numerical simulations show that the number of infected immune system cells in a vaccinated patient reaches approximately half of what is reached in a patient who undergoes natural infection. In turn, the level of immunity gained after one dose of vaccine is the same as the level observed in a measles patient 4 years after natural infection. The in-host model in \cite{Heffernan2008} was extended by the same authors to a population model (SEIRS) with waning immunity and immune system boosting \cite{Heffernan2009}. In contrast to classical SEIRS models, the class R refers here to individuals protected by short-term immune memory, while the class S refers to those individuals who have lost this short-term protection and may experience immune system boosting. Each compartment is classified according to the level of immunity, which can be related to the number of memory cells. Newborns are recruited into the susceptible class $S_0$ (lowest level of immunity). During exposure and infection the host does not change his level of immunity, that is, transition occurs from $S_j$ to $E_j$ to $I_j$ for each $j\in \N$. Hosts in $S$ and $R$ experience waning immunity and transit from $S_j$ to $S_{j-1}$ (respectively from $R_j$ to $R_{j-1}$). 
Immune system boosting is due to recovery from infection and is incorporated into the equations with transition terms from $I_j$ to $R_k$, with $k\geq j$. The resulting large system of ODEs, with a very high number of parameters, is quite hard to approach from an analytical point of view, hence the authors make use of numerical simulations to investigate the long term behavior. A somehow simplified version of the ODE system in \cite{Heffernan2009} was proposed by Reluga et~al. in \cite{Reluga2008}. A similar large system of ODEs was introduced by Lavine et~al. \cite{Lavine2011}, extending the SIRWS model in \cite{Mossong1999,Glass2003b}, by including several levels of immunity for immune hosts (R) and hosts with waning immunity (W), as well as age classes for all compartments. The authors claim that the model can explain several observed features of pertussis in US, in particular a shift in the age-specific incidence and the re-emergence of the disease in a highly vaccinated population.

\subsection{System of DDEs}
Delay models with constant or distributed delay have been introduced to describe waning of disease-induced or vaccine-induced immunity. A simple SIRS system with constant delay is given by
\begin{equation}
\label{sys:SIRSdelay}
\begin{aligned}
\dot S(t)&  = \mu(1-S(t))-\phi S(t)f(I(t))+\gamma I(t-\tau)e^{-\mu \tau}\\
\dot I(t)&  = \phi S(t)f(I(t))-(\mu+\gamma) I(t)\\
\dot R(t)& = \gamma I(t) -\mu R(t) -\gamma I(t-\tau)e^{-\mu \tau}.
\end{aligned}
\end{equation}
This model was studied by Kyrychko and Blyuss \cite{Kyrychko2005}, who provided results on existence, uniqueness and non-negativity of solutions, linear and global stability of the disease-free equilibrium, as well as global stability of the unique endemic equilibrium. A special case of \eqref{sys:SIRSdelay} was considered some years later by Taylor and Carr \cite{Taylor2009}. An extension of system \eqref{sys:SIRSdelay} with distributed delay was proposed in \cite{Blyuss2010} and shortly after in \cite{Bhat2012}.\\
\indent A more general model with distributed delay and vaccination was proposed by Arino et~al. in \cite{Arino2004}. Their system includes three compartments (susceptible, infective and vaccinated hosts) in a population which remains constant in time. Vaccine-induced immunity might be only partial, resulting in vaccinated individuals becoming infective.
Systems of ODEs or DDEs can be obtained from the general model by a proper choice of the kernel (see also \cite{Hethcote2000,Hethcote1981}).\\
\indent Recently, Yuan and B\'elair proposed a SEIRS model with integro-differential equations which resembles the systems in \cite{Arino2004,Hethcote2000}. The probability that an individual stays in the exposed class (E) for $t$ units of time is $P(t)$, hence,
\begin{displaymath}
E(t) = \int_0^t \beta \frac{S(u)I(u)}{N}e^{-b(t-u)}P(t-u)\,du.
\end{displaymath}
Similarly, $Q(t)$ is the probability that an individual is immune $t$ units of time after recovery, thus
\begin{displaymath}
R(t) = \int_0^t \gamma I(u)e^{-b(t-u)}Q(t-u)\,du.
\end{displaymath}
For a certain choice of the probabilities $P$ and $Q$, the problem can be reduced to a system with one or two constant delays.
The authors show existence of an endemic equilibrium and boundedness of solutions in a positive simplex. For the system with one constant delay, results for existence of a global attractor as well as the proof of persistence of the disease in case $R_0>1$ are provided.

\subsection{Systems of PDEs}
Structured populations in the context of waning immunity and immune system boosting have been motivated in different ways. Often the structure can be found in the biological age \cite{mclean1988,McLean1988a,Katzmann1984,Rouderfer1994}, and is used to observe disease transmission among babies, children, adults and seniors. Only few works suggest models for physiologically structured populations \cite{Martcheva2006,BarbarossaRostJoMB}.\\
\ \\
McLean and Anderson \cite{mclean1988,McLean1988a} proposed a model for measles transmission which includes a compartment for babies protected by maternal antibodies. Indeed, mothers who have had measles or have been vaccinated transfer measles immunity to the baby through the placenta. For several months after birth (ca. 2 months if the mother was vaccinated, ca. 4 months if she had the disease \cite{McLean1988a}) the baby is still protected by maternal antibodies and should not be vaccinated. The model by McLean and Anderson \cite{mclean1988} considers only waning of maternally induced immunity in the context of measles infection. Few years before McLean, Katzmann and Dietz \cite{Katzmann1984} proposed a bit more general model, which includes also waning of vaccine-induced immunity. In both cases, the age structure was used to determine the optimal age for vaccination. A compartment for adult hosts with waning immunity who can also receive immune system boosting was introduced only years later by Rouderfer et~al. \cite{Rouderfer1994}. A further deterministic system of ODEs for maternally induced immunity in measles was proposed in \cite{Moghadas2008}. \\
\ \\
Different is the approach when physiologically structured populations are considered. Martcheva and Pilyugin \cite{Martcheva2006} suggest an SIRS model in which infective and recovered hosts are structured by their immune status. In infective hosts the immune status increases over the course of infection, while in recovered hosts the immune status decays at some non-constant rate. When the immune status has reached a critical level, recovered hosts transit from the immune to the susceptible compartment.\\
\indent A general framework for SIRS systems, modeling waning immunity and immune system boosting, and combining the in-host perspective with the population dynamics, was proposed in \cite{BarbarossaRostJoMB}.

\section{A general modeling framework}
\label{sec:framework}
In this section we extend the model in \cite{BarbarossaRostJoMB} to include vaccine-induced immunity. As in \cite{Martcheva2006,BarbarossaRostJoMB}, we couple the in-host with the between-hosts dynamics, focusing on the effects of waning immunity and immune system boosting on the population dynamics. In contrast to the models proposed in \cite{Heffernan2009,Lavine2011}, we shall maintain the number of equations as low as possible. The resulting model (V1) is a system of  ODEs coupled with two PDEs. The ODE systems in \cite{Mossong1999,Glass2003b,Grenfell2012,Arino2006,Mossong2003}, as well as extensions of the DDEs systems in \cite{Taylor2009,Belair2013}, can be recovered from our modeling framework.\\
\ \\
Setting up our model we do not restrict ourselves to a particular pathogen. The model (V1) can be adapted to several epidemic outbreaks (e.g. measles, chickenpox, rubella, pertussis) by ad-hoc estimating coefficients from available experimental data \cite{Luo2012,Amanna2007,Li2013}.

\subsection{Model ingredients}
\subsubsection{\textit{Originally susceptible} and infectives hosts}
Let $S(t)$ denote the total population of \textit{originally susceptible} hosts. These are susceptible individuals which have neither received vaccination nor have been infected before. Newborns enter the susceptible population at rate $b(N)$, dependent on the total population size $N$. For simplicity we assume that the natural death rate $d>0$ does not depend on $N$. Assume that $b:[0,\infty)\to [0, b_+],\, N\mapsto b(N),$ with $0< b_+ <\infty$, is a nonnegative function, with $b(0)=0$. Finally, assume that in absence of disease-induced death there exists an equilibrium $N^*$ such that $b(N^*)=d\,N^*$.\\
\indent Let $I(t)$ denote the total infective population at time $t$. Infection of susceptible individuals occurs by contact, at rate $\beta I/N$. Infected hosts recover at rate $\gamma>0$. When we include disease-induced death at rate $d_I>0$, the equilibrium $N^*$ satisfies
\begin{equation*}
\label{eq:equil_Nstart_dI}
b(N^*)=d\,N^*+d_I I^*.
\end{equation*}

\subsubsection{Immune individuals}
Let us denote by $r(t,z)$ the density of recovered individuals with disease-induced immunity level $z \in [\zm, \zM]$ at time $t$. The total population of recovered hosts is given by 
\begin{displaymath}
R(t) = \int_{\zm}^{\zM} r(t,z)\, dz.
\end{displaymath}

\noindent The parameter $z$ describes the immune status and can be related to the number of specific immune cells of the host. The value $\zM$ corresponds to maximal immunity, whereas $\zm$ corresponds to low level of immunity. Individuals who recover at time $t$ enter the immune compartment with maximal level of immunity $\zM$. The level of immunity tends to decay in time and when it reaches the minimal value $\zm$, the host becomes susceptible again. However, exposure to the pathogen can boost the immune system from $z\in [\zm,\zM]$ to any higher status. It is not straightforward to determine how this kind of immune system boosting works, as no experimental data are available. Nevertheless, laboratory analysis on vaccines tested on animals or humans suggest that the boosting efficacy might depend on several factors, among which the current immune status of the recovered host and the amount of pathogen he receives \cite{Amanna2007,Luo2012}. Possibly, exposure to the pathogen can restore the maximal level of immunity, just as natural infection does \cite{BarbarossaRostJoMB}.\\
\indent Let $p(z,\tilde z),\, z\geq \tilde z,\,z,\tilde z \in \R$ denote the probability that an individual with immunity level $\tilde z$ moves to immunity level $z$, when exposed to the pathogen. Due to the definition of $p(z,\tilde z)$, we have 
$p(z,\tilde z)\in [0,1],\, z\geq \tilde z$ and 
\begin{displaymath}
p(z,\tilde z)= 0, \quad \mbox{for all} \quad z < \tilde z.
\end{displaymath}
As we effectively consider only immunity levels in the interval $[\zm,\zM]$, we set
\begin{displaymath}
p(z,\tilde z)= 0, \quad \mbox{for all} \quad \tilde z \in (-\infty,\zm) \cup (\zM,\infty).	
\end{displaymath}
Then we have
$$\int_{-\infty}^{\infty}p(z,\tilde z)\, dz\,=\,\int_{\tilde z}^{\zM}p(z,\tilde z)\, dz\,=\,1,\quad  \mbox{for all}\quad \tilde z \in [\zm,\zM].$$
Exposure to the pathogen might restore exactly the immunity level induced by the disease ($\zM$). In order to capture this particular aspect of immune system boosting, we write the probability $p(z,\tilde z)$ as the combination of a continuous ($p_0$) and atomic measures (Dirac delta):
\begin{displaymath}
p(z,\tilde z)= c_{max}(\tilde z)\delta (\zM-\tilde z) + c_0(\tilde z)p_0(z,\tilde z) + c_1(\tilde z)\delta(z-\tilde z),
\end{displaymath}
where 
\begin{itemize}
  \item \textbf{$c_{max}:[\zm,\zM ]\to [0,1],\;y\mapsto c_{max}(y)$}, is a continuously differentiable function and describes the probability that, due to contact with infectives, a host with immunity level $y$ boosts to the maximal level of immunity $\zM$.
	\item \textbf{$c_{0}:[\zm,\zM]\to [0,1],\;y\mapsto c_{0}(y)$}, is a continuously differentiable function and describes the probability that, due to contact with infectives, a host with immunity level $y$ boosts to any other level $z \in (y,\zM)$, according to the continuous probability $p_0(z,y)$.
	\item \textbf{$c_{1}(y)=1-c_{max}(y)-c_0(y)$} describes the probability that getting in contact with infectives, the host with immunity level $y\in [\zm,\zM]$ does not experience immune system boosting.
	\end{itemize}

\indent The immunity level decays in time at some rate $g(z)$ which is the same for all recovered individuals with immunity level $z$. In other words, the immunity level $z$ follows
\begin{equation*}
\frac{d}{dt}z(t)=g(z),
\end{equation*}
with $g:[\zm,\zM]\to (0,K_g],\; K_g<\infty$ continuously differentiable. The positivity of $g(z)$ is required from the biological motivation. Indeed, if  $g(\tilde z)=0$ for some value $\tilde z \in [\zm,\zM]$, there would be no change of the immunity level at $\tilde z$, contradicting the hypothesis of natural decay of immune status. In absence of immune system boosting, we have that
$$ \int_{\zm}^{\zM} \frac{1}{g(x)}\,dx$$
is the time a recovered host remains immune (see \cite{BarbarossaRostJoMB}).

\subsubsection{Vaccination}
We structure the vaccinated population by the level of immunity as well. Let $v(t,z)$ be the density of vaccinees with immunity level $z \in [\zm, \zM]$ at time $t$. The total population of vaccinated hosts is given by 
\begin{displaymath}
V(t) = \int_{\zm}^{\zM} v(t,z)\, dz.
\end{displaymath}
Vaccination infers a level of immunity $\zV,$ which is lower than the level of immunity after natural infection: $\zM>\zV>\zm$ \cite{Siegrist2008}. As in recovered individuals, the level of immunity of a vaccinated host tends to decay in time and when it reaches the minimal value $\zm$, the host becomes susceptible again. However, also in vaccinated hosts, exposure to the pathogen can boost the immunity level $z\in [\zm,\zV]$ to any higher value in $ [\zm,\zM]$. Immune system boosting is described by the probability $p(z,\tilde z)$, as in recovered hosts. We consider the possibility that exposure to the pathogen boosts the immune system of a vaccinated individual to $z \in (\zV,\zM]$. Vaccinated hosts with $z \in (\zV,\zM]$ have an immune status which can be compared to the one of hosts who recovered from natural infection.\\
\indent It is reasonable to assume that in vaccinated individuals the immunity level decays in time at the same rate $g$, as in hosts who underwent natural infection. In absence of exposure to the pathogen (hence in absence of immune system boosting), the time that a vaccinee remains immune is shorter than the time a recovered host does:
$$ \int_{\zm}^{\zV} \frac{1}{g(x)}\,dx<\int_{\zm}^{\zM} \frac{1}{g(x)}\,dx.$$
Let us define the vaccination rate at birth $\alpha>0$. We assume that originally susceptible (adult) individuals get vaccinated at rate $\phi\geq 0$. 

\subsubsection{Becoming susceptible again}
In absence of immune system boosting both disease-induced and vaccine-induced immunity fade away. Individuals who lose immunity either after recovery from infection or after vaccination, enter the class $S_2$ of susceptible individuals who shall not get a new dose of vaccine. A host who had the disease or got vaccination relies indeed on the induced-immunity and is not aware of the fact that his level of immunity might have dropped below the critical immunity threshold.\\
\ \\
We denote by $S_2(t)$ the population at time $t$ of susceptible hosts who are not going to receive vaccination.

\subsection{Model equations}
In view of all what we have mentioned above, we can easily write down the equations for the compartments $S,\,I$ and $S_2$. Let initial values $S(0)=S^0\geq 0$, $I(0)=I^0\geq 0$ and $S_2(0)=S_2^0\geq 0$ be given. The population of originally susceptible individuals is governed by  
\begin{equation}
\dot S(t) = \underbrace{b(N(t))(1-\alpha)}_{\mbox{birth}} -\underbrace{\phi S(t)}_{\mbox{vaccination}}-\underbrace{\beta \frac{S(t)I(t)}{N(t)}}_{\mbox{infection}}-\underbrace{dS(t)}_{\mbox{death}},
\label{eq:S_vacc}
\end{equation}
whereas hosts who become susceptible due to immunity loss follow 
\begin{equation*}
\dot S_2(t) = -\underbrace{\beta \frac{S_2(t)I(t)}{N(t)}}_{\mbox{infection}}-\underbrace{dS_2(t)}_{\mbox{death}}
 +\underbrace{\Lambda_R}_{\substack{\text{immunity loss}\\ \text{after recovery}}}+\underbrace{\Lambda_V}_{\substack{\text{immunity loss}\\ \text{after vaccination}}}.
\end{equation*}
The term $\Lambda_R$ (respectively $\Lambda_V$), which represents transitions from the immune (respectively, the vaccinated) compartment to the susceptible one, will be specified below together with the dynamics of the recovered (respectively, vaccinated) population.\\
\ \\
Both kinds of susceptible hosts can become infective due to contact with infective hosts:
\begin{equation}
\dot I(t) = \underbrace{\beta \frac{S(t)I(t)}{N(t)}}_{\mbox{infection of }S}+\underbrace{\beta \frac{S_2(t)I(t)}{N(t)}}_{\mbox{infection of }S_2} -\underbrace{\gamma I(t)}_{\mbox{recovery}} -\underbrace{d I(t)}_{\substack{\text{natural}\\ \text{death}}}-\underbrace{d_I I(t)}_{\substack{\text{disease-induced}\\ \text{death}}}.
\label{eq:I}
\end{equation}
To obtain an equation for the recovered individuals, structured by their levels of immunity, one can proceed similarly to size structured models or as it was done for the immune population in \cite{BarbarossaRostJoMB}. The result is the following PDE. Let a nonnegative initial distribution $r(0,z)=\psi(z),\, z \in  [\zm,\zM]$ be given. For $t>0, z\in[\zm,\zM]$ we have
\begin{equation}
\begin{aligned}
\frac{\partial }{\partial t}r(t,z)-\frac{\partial }{\partial z}\left(g(z)r(t,z)\right) & = 
-dr(t,z)+ \beta \frac{I(t)}{N(t)}\int_{\zm}^{z} p(z,x)r(t,x)\,dx\\[0.5em]
& \phantom{==} - r(t,z)\beta\frac{I(t)}{N(t)},
\label{eq:pdeR_zmzV}
\end{aligned}
\end{equation}
with the boundary condition 
\begin{equation}
\label{eq:BC_mod1_pde_R} 
g(\zM)r(t,\zM)  = \gamma I(t) + \beta \frac{I(t)}{N(t)}\int_{\zm}^{\zM} p(\zM,x)r(t,x)\,dx.
\end{equation}
Equation \eqref{eq:pdeR_zmzV} expresses the rate of change in the density of recovered individuals according to immune level due to natural waning, mortality, and boosting. The boundary condition \eqref{eq:BC_mod1_pde_R} includes newly recovered individuals as well as those recovered individuals, who just received a boost which elevated their immune system to maximal level.

Next we shall consider the vaccinated population. Again, by structuring this group according to immunity level, one has the PDE
\begin{equation}
\begin{aligned}
\frac{\partial }{\partial t}v(t,z) & = \frac{\partial }{\partial z}\left(g(z)v(t,z)\right)  
-dv(t,z)+ \beta \frac{I(t)}{N(t)}\int_{\zm}^{z} p(z,x)v(t,x)\,dx\\[0.5em]
& \phantom{==} - v(t,z)\beta\frac{I(t)}{N(t)}+\delta(z-z_{vax})\left(\phi S(t) + \alpha b(N(t))\right),
\label{eq:mod1_pde_V} 
\end{aligned}
\end{equation}
and
\begin{equation}
\label{eq:BC_mod1_pde_V} 
g(\zM)v(t,\zM)  = \beta \frac{I(t)}{N(t)}\int_{\zm}^{\zM} p(\zM,x)v(t,x)\,dx,
\end{equation}
provided with a nonnegative initial distribution $v(0,z)=\psi_v(z),\, z \in  [\zm,\zM]$. Observe that newly vaccinated hosts do not enter the vaccinated population at $\zM$, but at the lower value $\zV$, which is expressed in equation \eqref{eq:mod1_pde_V} as an impulse at $z=z_{vax}$ by the term with the Dirac delta $\delta(z-z_{vax})$.

It becomes evident that the quantity $\Lambda_R$, initially introduced in the $S_2$ equation to represent the number of hosts who experienced immunity loss, is given by the number $g(\zm)r(t,\zm)$  of immune hosts who reached the minimal level of immunity after recovery from natural infection. Similarly, $\Lambda_V$ is the number $g(\zm)v(t,\zm)$ of vaccinated hosts who reached the minimal level of immunity. Hence we have
\begin{equation}
\dot S_2(t) = -\underbrace{\beta \frac{S_2(t)I(t)}{N(t)}}_{\mbox{infection}}-\underbrace{dS_2(t)}_{\mbox{death}}
 +\underbrace{g(\zm)r(t,\zm)}_{\Lambda_R}+\underbrace{g(\zm)v(t,\zm)}_{\Lambda_V}.
\label{eq:S2_vacc}
\end{equation}
\noindent In the following we refer to the complete system \eqref{eq:S_vacc} -- \eqref{eq:S2_vacc} as to \textbf{model (V1)}.

\section{Connection to other mathematical models}

\subsection{Connection to ODE models}
As it was shown in \cite{BarbarossaRostJoMB} for a simpler problem, model (V1) can be reduced to a system of ODEs analogous to those proposed in \cite{Mossong1999,Glass2003b,Grenfell2012,Lavine2011,Heffernan2009,Mossong2003}. The connection between model (V1) and the ODE system is given by the \textit{method of lines}, a technique in which all but one dimensions are discretized \cite{MOLbook}. In our case, we shall discretize the immunity level ($z$) and obtain a system of ODEs in the time variable.\\
\ \\
Let us define a sequence $\left\{z_j\right\}_{j\in \N}$, with $h_j:=z_{j+1}-z_j>0$, for all $j \in \N$. To keep the demonstration as simple as possible, we choose a grid with only a few points, $z_1:=\zm<\zw:=\zV<\zf<\zM$ and for simplicity (or possibly after a rescaling) assume that $h_j=1$ for all $j$. We define the following subclasses of the immune/vaccinated population:
\begin{itemize}\itemsep0.5cm
	\item $R_F(t):= r(t,\zf)$, immune hosts with high level of immunity at time $t$. As their immunity level is quite high, these individuals do not experience immune system boosting. Immunity level decays at rate $\mu:=g(\zf)>0$.
	\item $R_W(t):= r(t,\zw)$, immune hosts with intermediate level of immunity at time $t$. These individuals can get immune system boosting and move to $R_F$. Immunity level decays at rate $\nu:=g(\zw)>0$.
	\item $R_C(t):= r(t,\zm)$, immune hosts with critically low level of immunity at time $t$. With probability $\theta$ boosting moves $R_C$ individuals to $R_W$ (respectively, with probability $(1-\theta)$ to $R_F$). Immunity level decays at rate $\sigma:=g(\zm)>0$. If they do not get immune system boosting, these hosts move to the class $S_2$ (become susceptible again). 
		\item $V_R(t):= v(t,\zf)$, vaccinated hosts who thanks to immune system boosting gained a very high level of immunity at time $t$. These individuals do not experience immune system boosting. Immunity level decays at rate $\mu$.
	\item $V_0(t):= v(t,\zw)$, vaccinated individuals at time $t$ with maximal vaccine-induced immunity. This class includes new vaccinees. If their immune system gets boosted hosts move to $V_R$. Immunity level decays at rate $\nu$.
	\item $V_C(t):= v(t,\zm)$, vaccinees with critically low level of immunity at time $t$. With probability $\xi$ boosting moves $V_C$ hosts to $V_0$ and with probability $(1-\xi)$ to $V_R$. Immunity level decays at rate $\sigma$. If they do not receive immune system boosting, $V_C$ hosts move to $S_2$.	
	\end{itemize}
To show how the PDE system can be reduced to a system of ODEs by means of the method of lines, we consider a simple example. Let us neglect immune system boosting for a moment. Then the PDE for $r(t,z)$ in model (V1) becomes
\begin{equation}
\label{eq:PDEr_MOL_noboost}
\frac{\partial}{\partial t} r(t,z) = \frac{\partial}{\partial z} \bigl(g(z) r(t,z)\bigr) -d r(t,z), \qquad z \in [\zm,\zM],
\end{equation}
with boundary condition $R_{\zM}(t):=r(t,\zM)=\gamma I(t) / g(\zM)$. 
Using forward approximation for the $z$-derivative in \eqref{eq:PDEr_MOL_noboost}, we obtain, e.g., for $R_F(t)$ the following differential equation:
\begin{align*}
\dot R_F(t) & =\frac{\partial}{\partial t} r(t,\zf)\\
            & = \frac{\partial}{\partial z} \bigl(g(\zf) r(t,\zf)\bigr) -d r(t,\zf)\\
						& \approx \frac{g(\zM) r(t,\zM)- g(\zf) r(t,\zf)}{\underbrace{\zM-\zf}_{=1}} -d r(t,\zf)\\
						& = g(\zM) R_{\zM}(t)- \mu R_F(t) -d R_F(t)\\
						& = \gamma I(t)- (\mu+d) R_F(t).
\end{align*}
Analogously one can find equations for $R_W,\,R_C,\,V_R,\,V_0$ and $V_C$. Altogether we obtain a system of ordinary differential equations in which a linear chain of ODEs replaces the PDEs for the immune and the vaccinated class:
\begin{align*}
\dot S(t) & = (1-\alpha)b(N(t)) -\phi S(t)-\beta \frac{S(t)I(t)}{N(t)}-dS(t)\\
\dot I(t) & = \beta \frac{I(t)}{N(t)}(S(t)+S_2(t))-(\gamma+d+d_I)I(t)\\
\dot R_F(t) & = \gamma I(t)-\mu R_F(t) -dR_F(t)\\
\dot R_W(t) & =\mu R_F(t)-\nu R_W(t)-d R_W(t)\\
\dot R_C(t) & = \nu R_W(t)-\sigma R_C(t) -d R_C(t)\\
\dot V_R(t) & = -\mu V_R(t) -d V_R(t)\\
\dot V_0(t) & = \phi S(t)+\alpha b (N(t)) +\mu V_R(t)-\nu V_0(t) -dV_0(t)\\
\dot V_C(t) & =\nu V_0(t)-\sigma V_C(t) -d V_C(t)\\
\dot S_2(t) &= -\beta \frac{S_2(t)I(t)}{N(t)}-dS_2(t)+\sigma (R_C(t)+ V_C(t)).
\end{align*}
The method of lines can be applied to the full model (V1) as well \cite{BarbarossaRostJoMB}. To this purpose it is necessary to discretize the boosting probability $p(z,\tilde z)$ (this is expressed by the parameters $\xi$ and $\theta$ below). Incorporating the boosting effect, the result is the following system of ODEs.

\begin{align*}
\dot S(t) & = (1-\alpha)b(N(t)) -\phi S(t)-\beta \frac{S(t)I(t)}{N(t)}-dS(t)\\
\dot I(t) & = \beta \frac{I(t)}{N(t)}\left(S(t)+S_2(t)\right)-(\gamma+d+d_I)I(t)\\
\dot R_F(t) & = \gamma I(t)-\mu R_F(t) -dR_F(t)+\beta \frac{I(t)}{N(t)} \left(R_W(t) +(1-\theta)R_C(t)\right)\\
\dot R_W(t) & =\mu R_F(t)-\nu R_W(t)-d R_W(t)+ \beta \frac{I(t)}{N(t)}(\theta R_C(t)-R_W(t))\\
\dot R_C(t) & = \nu R_W(t)-\sigma R_C(t) -d R_C(t)-\beta \frac{I(t)}{N(t)} R_C(t)\\
\dot V_R(t) & = \beta \frac{I(t)}{N(t)} \left(V_0(t) +(1-\xi)V_C(t) \right)-\mu V_R(t) -dV_R(t)\\
\dot V_0(t) & = \phi S(t)+\alpha b (N(t)) +\mu V_R(t)-\nu V_0(t) -dV_0(t) +\beta \frac{I(t)}{N(t)}\left(\xi V_C(t)-V_0(t)\right)\\
\dot V_C(t) & = \nu V_0(t)-\sigma V_C(t) -d V_C(t)-\beta \frac{V_C(t)I(t)}{N(t)}\\
\dot S_2(t) &= -\beta \frac{S_2(t)I(t)}{N(t)}-dS_2(t)+\sigma (R_C(t)+ V_C(t)).
\end{align*}
The linear chain of ODEs provides a rough approximation of the PDEs in model (V1). Indeed, with the method of lines we approximate the PDE dynamics considering only changes at the grid points ($\zm,\;\zw,\;\zf$), whereas the dynamics remains unchanged in each immunity interval $[z_{j},z_{j+1}]$. We consider as representative point of the interval the lowest boundary $z_j$ - for this reason we do not have a differential equation for $R_{\zM}(t)$ or $V_{\zM}(t)$.

\subsection{Connection to DDE models}
\label{sec:connDDEs}
Delay models with constant delay can be recovered from special cases of model (V1). We show here how to obtain the classical SIRS model with delay studied in \cite{Taylor2009}, or extensions thereof.\\
\ \\ 
In the following we neglect boosting effects and vaccination. Further we do not distinguish between originally susceptibles and host who have lost immunity, hence w.r.t. model (V1) we identify the classes $S$ and $S_2$. From our assumptions, the disease-induced immunity lasts for a fix time, $\tau>0$ years, given by
$$ \int_{\zm}^{\zM}\frac{1}{g(x)}\,dx=\tau.$$
We can express the total immune population at time $t$ as the number of individuals who recovered in the time interval $[t-\tau,t]$, 
\begin{equation*}
R(t) = \gamma \int_{t-\tau}^{t} I(y)e^{-d(t-y)}\,dy = \gamma \int_0^{\tau} I(t-x)e^{-dx}\,dx.
\label{eq:defR_integr}
\end{equation*}
Differentiation with respect to $t$ yields
\begin{equation}
\dot R(t) = \gamma I(t)-\gamma I(t-\tau)e^{-d\tau}-dR(t).
\label{eq:ddeR1}
\end{equation}
On the other side, we have the definition in terms of distribution of immune individuals,
\begin{equation*}
R(t) =\int_{\zm}^{\zM} r(t,z)\,dz.
\end{equation*}
Differentiate the last relation and compare with \eqref{eq:ddeR1}:
\begin{equation*}
g(\zM)r(t,\zM)=\gamma I(t), \qquad g(\zm)r(t,\zm)=\gamma I(t-\tau)e^{-d\tau}.
\label{eq:ddeR_relations}
\end{equation*}
This means that individuals with maximal level of immunity are those who recover from infection. If a host who recovers at time $t_1$ survives up to time $t_1+\tau$, he exits the $R$ class and enter $S$. 
In turn, we find a delay term in the equation for $S$ too, and have a classical SIRS model with constant delay 
\begin{align*}
\dot S(t) & = b(N(t)) -\beta \frac{S(t)I(t)}{N(t)}-dS(t)+\gamma I(t-\tau)e^{-d\tau}\\[0.3em]
\dot I(t) & = \beta \frac{S(t)I(t)}{N(t)}-(\gamma+d+d_I)I(t)\\[0.3em]
\dot R(t) & =\gamma I(t)-\gamma I(t-\tau)e^{-d\tau} -dR(t),
\end{align*}
which was studied by Taylor and Carr \cite{Taylor2009}.\\
\ \\
Now we can include again vaccination and the class $S_2$ as in the general model (V1). We assume that vaccine-induced immunity lasts for a time $\tau_v>0$,
$$ \tau_v:=\int_{\zm}^{\zV}\frac{1}{g(x)}\,dx\;<\;\int_{\zm}^{\zM}\frac{1}{g(x)}\,dx=:\tau.$$
With similar arguments as for the immune population, we obtain the relations
\begin{align*}
g(\zV)v(t,\zV) & =\alpha b(N(t))+\phi S(t), \\
g(\zm)v(t,\zm) & =\left(\alpha b(N(t-\tau_v))+\phi S(t-\tau_v)\right)e^{-d\tau_v},
\end{align*}
and find a system with two constant delays
\begin{equation*}
\begin{aligned}
\dot S(t) & = (1-\alpha)b(N(t)) -\phi S(t) -\beta \frac{S(t)I(t)}{N(t)}-dS(t)\\[0.3em]
\dot I(t) & = \beta \frac{I(t)}{N(t)}(S(t)+S_2(t))-(\gamma+d+d_I)I(t)\\[0.3em]
\dot R(t) & =\gamma I(t)-\gamma I(t-\tau)e^{-d\tau} -dR(t)\\[0.3em]
\dot V(t) & = \alpha b(N(t))+\phi S(t)-\left(\alpha b(N(t-\tau_v))+\phi S(t-\tau_v)\right)e^{-d\tau_v}  -d V(t)\\
\dot S_2(t) &= -\beta \frac{S_2(t)I(t)}{N(t)}-dS_2(t)+\gamma I(t-\tau)e^{-d\tau}\\
 & \phantom{=} +\left(\alpha b(N(t-\tau_v))+\phi S(t-\tau_v)\right)e^{-d\tau_v}.
\end{aligned}
\end{equation*}

\section*{Acknowledgments}
Authors were supported by the ERC Starting Grant No 259559. MVB was supported by the European Union and the State
of Hungary, co-financed by the European Social Fund in the framework of T\'AMOP-4.2.4.
A/2-11-1-2012-0001 National Excellence Program. GR was supported by Hungarian Scientific Research Fund OTKA K109782 and T\'AMOP-4.2.2.A-11/1/KONV-2012-0073 "Telemedicine focused
research activities on the field of Mathematics, Informatics and Medical
sciences".

\end{document}